\documentstyle[12pt,epsfig]{article}
\begin{document}

\begin{titlepage}
\rightline{\large August 2003}
\rightline{\large hep-ph/0308254}
\vskip 2cm
\centerline{\Large \bf Implications of the DAMA and CRESST experiments 
}
\vskip 0.2cm
\centerline{\Large \bf for mirror matter-type dark matter}
\vskip 2.2cm
\centerline{\large R. Foot\footnote{
E-mail address: rfoot@unimelb.edu.au}}

\vskip 0.7cm
\centerline{\large \it School of Physics,}
\centerline{\large \it University of Melbourne,}
\centerline{\large \it Victoria 3010 Australia}
\vskip 2cm
\noindent
Mirror atoms are expected to be a significant component of the galactic
dark matter halo if mirror matter is identified 
with the non-baryonic dark matter in
the Universe. Mirror matter can interact with ordinary matter
via gravity and via the photon-mirror photon kinetic mixing
interaction -- causing  mirror charged particles to couple
to ordinary photons with effective electric charge $\epsilon e$.
This means that the nuclei of mirror atoms can elastically
scatter off the nuclei of ordinary atoms, leading to 
nuclear recoils, which can be detected in existing dark matter experiments.
We show that the dark matter experiments
most sensitive to this type of dark matter candidate (via
the nuclear recoil signature) are the
DAMA/NaI and CRESST/Sapphire experiments.
Furthermore, we show that the impressive annual modulation signal
obtained by the DAMA/NaI experiment can be explained by mirror
matter-type dark matter
for $|\epsilon | \sim 5 \times 10^{-9}$ and is 
supported by DAMA's absolute rate measurement as well
as the CRESST/Sapphire data. This value of 
$|\epsilon |$ is consistent with the value obtained from
various solar system anomalies including the Pioneer spacecraft
anomaly, anomalous meteorite events and lack of small
craters on the asteroid Eros. It is also consistent with
standard BBN.

\end{titlepage}

The DAMA/NaI experiment\cite{dama,dama2} has been searching for dark 
matter and has obtained
some very exciting positive results which merit serious
consideration.
While they have interpreted their data in terms of weakly 
interacting heavy particles 
an alternative interpretation will be suggested here.

In the DAMA/NaI experiment the target consists of 100 kg of 
radiopure NaI. The aim of the experiment is to measure
recoil energy of the Na, I atoms due to interactions of
dark matter particles with their detector. 
Due to the Earth's motion around the sun, 
the rate should experience a small annual
modulation:
\begin{eqnarray}
A\cos 2\pi (t - t_0)/T
\label{1}
\end{eqnarray}
According to the DAMA analysis\cite{dama2}, they indeed find such a modulation
over 7 annual cycles at more than $6\sigma$ C.L. Their data fit 
gives $T = (1.00 \pm 0.01)$ year and 
$t_0 = 144 \pm 22$ days, consistent with the expected
values. [The expected value for $t_0$ is 152 days (2 June), where the
Earth's velocity, $v_E$, reaches
a maximum with respect to the galaxy].  
The strength of their signal is $A = (0.019 \pm 0.003)$
cpd/kg/keV.

The DAMA collaboration have interpreted these impressive results as
evidence for heavy weakly interacting dark matter 
particles. However, another possibility is that this
experiment has observed the impacts of galactic mirror atoms,
as will shortly be explained.

Mirror matter is predicted to exist if nature exhibits
an exact unbroken mirror symmetry\cite{flv} (for
reviews and more complete set of references, see Ref.\cite{review}). 
For each type of ordinary
particle (electron, quark, photon etc) there is a mirror partner
(mirror electron, mirror quark, mirror photon etc), 
of the same mass. The two sets of particles form 
parallel sectors each with gauge symmetry $G$
(where $G = SU(3) \otimes SU(2) \otimes U(1)$ in the 
simplest case)
so that the full gauge group is $G \otimes G$.
The unbroken mirror symmetry maps
$x \to -x$ as well as ordinary particles into mirror
particles. Exact unbroken time reversal symmetry
also exists, with standard CPT identified as the product
of exact T and exact P\cite{flv}.

Ordinary and mirror particles can interact with each
other by gravity and via the photon-mirror
photon kinetic mixing interaction:
\begin{eqnarray}
{\cal{L}} = {\epsilon \over 2} F^{\mu \nu} F'_{\mu \nu} 
\end{eqnarray}
where $F^{\mu \nu}$
($F'_{\mu \nu}$) is the field strength tensor for electromagnetism
(mirror electromagnetism)
\footnote{Given the 
constraints of gauge invariance, renomalizability and mirror
symmetry it turns out\cite{flv} 
that the only allowed non-gravitational interactions
connecting the ordinary particles with the mirror particles
are via photon-mirror photon kinetic mixing 
and via a Higgs-mirror Higgs quartic
interaction, ${\cal{L}} = \lambda \phi^{\dagger} \phi \phi'^{\dagger}
\phi'$. If neutrinos have mass, then ordinary - mirror
neutrino oscillations may also occur\cite{flv2,f}.}. 
Photon-mirror photon mixing causes 
mirror charged particles to couple to 
ordinary photons with a small effective
electric charge, $\epsilon e$\cite{flv,hol,sasha}.
Interestingly, the existence
of photon-mirror photon kinetic mixing
allows mirror matter to explain a number of puzzling
observations, including the Pioneer spacecraft anomaly\cite{p1,p2},
anomalous meteorite events\cite{fy,doc}
and the unexpectedly low number of small craters on the
asteroid 433 Eros\cite{fm,eros}. It turns out that these 
explanations and other constraints\cite{fg,ortho} suggest that
$\epsilon$ is in the range 
\begin{eqnarray}
10^{-9} \stackrel{<}{\sim} |\epsilon | \stackrel{<}{\sim} 
5\times 10^{-7}.
\label{range}
\end{eqnarray}

More generally, mirror matter is a rather obvious candidate 
for the non-baryonic dark matter in the Universe because:
\begin{itemize}
\item
It is well motivated from fundamental physics
since it is required to exist if parity and time reversal
symmetries are exact, unbroken symmetries of nature.
\item
It is necessarily dark and stable. Mirror baryons have
the same lifetime as ordinary baryons and couple to mirror
photons instead of ordinary photons.
\item
Mirror matter can provide a suitable framework for
which to understand the large scale structure of the 
Universe\cite{comelli}.
\item
Recent observations from WMAP\cite{wmap} and other experiments suggest
that the cosmic abundance of non-baryonic dark matter is
of the same order of magnitude as ordinary matter $\Omega_b 
\sim \Omega_{dark}$. A result which can naturally occur if
dark matter is identified with mirror matter\cite{fvwmap}.
\end{itemize}

If mirror matter is identified as the non-baryonic dark matter,
then the dark matter halo will consist of compact objects
such as mirror stars and planets, as well as a mirror gas and
dust component. Evidence for mirror stars arises from 
MACHO observations \cite{macho,macho2}
(and to some extent from the puzzling `isolated'
planets\cite{isol}) while the existence of close-in extrasolar planets
can also be viewed as mirror matter manifestations\cite{exo}.
The amount of material
in compact form is probably less than $50\%$ (coming
from the MACHO upper limit). Thus we expect a dark matter
halo with a significant gas/dust component containing mirror
$H', He'$ + heavier mirror elements. 
Assuming a local halo dark matter energy density of $0.3 \ GeV/cm^3$, then the
number densities of $A' = H', He'$ and heavier elements is then given by
\begin{eqnarray}
n_{A'} = \xi_{A'} {0.3 \ GeV \over M_{A'}}  \  cm^{-3}
\label{08}
\end{eqnarray}
where $\xi_{A'} \equiv \rho_{A'}/(0.3 \ {\rm GeV/cm^3})$ 
is the $A'$ proportion (by mass) of the halo dark matter.
As discussed above, a plausible value for $\sum_{A'} \xi_{A'}$ is 
$ \sim 1/2$.

Arguments from early Universe cosmology (mirror
BBN)\cite{comelli} suggest that $He'$ dominates over $H'$, quite
unlike the case with ordinary matter. 
Mirror elements heavier than $H', He'$ will presumably 
come from nucleosynthesis within mirror stars, qualitatively
similar to the ordinary matter case. In the ordinary
matter case,
the galactic relative (mass) abundance of elements heavier
than $H, He$ (collectively called `metals' in the
astrophysics literature) is estimated\cite{wh} to be roughly $Z_g \sim 0.02 
\ \Rightarrow \ \xi_{Metals}/\xi_{He} \sim 0.10$.
These heavier elements are made up primarily ($> 90\%$) of
O, Ne, N, C which have $M_A/M_P \simeq 16 \pm 4,\ Z = 8 \pm 2$.
Thus, oxygen provides an excellent `average' for ordinary elements
heavier than helium -- except perhaps for iron
[which is about 10 times less abundant (by mass) than oxygen].
In the case of mirror element abundances, we would expect a
{\it qualitatively} similar picture, i.e. $O'$ (and
elements with nearby atomic number) should dominate the energy
density after $H', He'$, with a possible small $Fe'$ contribution.
Thus we need only consider four mirror
elements: $H', He', O', Fe'$ (where $O'$ stands for Oxygen
and nearby elements).
Of course, {\it quantitatively},
the ratios $\xi_{O'}/\xi_{He'}, \xi_{Fe'}/\xi_{O'}$ are quite uncertain
because of the different initial values for $He'/H'$
(coming from mirror BBN) and other different initial conditions.
Although the proportion of the various mirror elements in the
halo (gas/dust ratio etc) is uncertain, 
the mass scale is not a free parameter:
mirror hydrogen, $H'$, is predicted to have exactly the same mass as ordinary
hydrogen, i.e. $M_{H'} = 0.94\ GeV$, mirror helium, $He'$, has mass
$M_{He'} = 3.76\ GeV$ etc\footnote{
It is possible to construct mirror matter models with broken
mirror symmetry, in which case the masses of the mirror
particles need not be the same as their ordinary counterparts\cite{broken}, 
but these models tend to be more complicated
and/or less well motivated than the simplest case of
unbroken mirror symmetry\cite{com}.}.

In an experiment such as DAMA/NaI, the measured quantity is 
the recoil energy, $E_R$, of a target atom. The minimum velocity
of a mirror atom of mass $M_{A'}$ impacting on
a target atom of mass $M_A$ is related to $E_R$ via 
the kinematic relation:
\begin{eqnarray}
v_{min} &=& \sqrt{ {(M_A + M_{A'})^2 E_R\over 2M_A M^2_{A'}} } .
\label{v}
\end{eqnarray}
Interestingly, most of the existing dark matter experiments are not very
sensitive to mirror matter-type dark matter because $v_{min}$ [Eq.(\ref{v})]
turns out to be too high. This is because they either use target
elements which are too heavy (i.e. large $M_A$) or have a 
$E_R$ threshold which is too high.
For example, the CDMS experiment uses $Ge$ as the target
material and has a threshold of 10 keV\cite{cdms}. This means
that $v_{min} \approx 1600$ km/s (for $He'$).
Although there would be no cutoff velocity at the galactic escape
velocity for $He'$ due to $He'$ self interactions, 
the number of $He'$ with such high velocities would
be negligible.
The existing experiments with the greatest sensitivity 
to light mirror elements are the DAMA/NaI\cite{dama,dama2} and
the CRESST/sapphire experiments\cite{cresst}. Both of these
experiments will be examined in detail.

When a mirror atom
(of mass $M_{A'}$, atomic number $Z'$) 
encounters ordinary matter
(comprised of atoms with mass $M_A$, atomic number $Z$) Rutherford
scattering can occur, 
with center of mass cross section:
\footnote{
Note that unless otherwise stated, we use natural units where
$\hbar = c = 1$.}
\begin{eqnarray}
\left(\frac{d\sigma}{d\Omega}\right)_{\rm{elastic}} = 
\frac{\epsilon^2 \alpha^2 Z^2 Z'^2 M^2_{red}}{
4M_{A'}^4 v^4_{cm} \sin^4 \frac{\theta_s}{2} }
\ F^2_A (qr_{A}) F^2_{A'}(qr_{A'})
\label{mon2}
\end{eqnarray}
where $v_{cm}$ is the center of mass velocity of
the impacting mirror atom and 
$M_{red} = M_A M_{A'}/(M_A + M_{A'})$ is the reduced mass
\footnote{Due to the screening effects of the atomic electrons,
the cross section is modified (and becomes suppressed)
at small scattering angles [$\theta_s
\stackrel{<}{\sim} 1/(M_{A'} v r_0)$ with $r_0 \sim 10^{-9}$ cm].}. 
In Eq.(\ref{mon2}), $F_{X} (qr_X)$ ($X = A, A'$) are the
form factors which take into account
the finite size of the nuclei and mirror nuclei.
($q = (2M_A E_R)^{1/2}$ is the momentum transfer and
$r_X$ is the effective nuclear radius). 
A simple analytic expression for the form factor, which we
adopt in our numerical work, is\cite{ls}:
\begin{eqnarray}
F_X (qr_X) = 3{j_1 (qr_X) \over qr_X} \times e^{-(qs)^2/2}
\label{ff}
\end{eqnarray}
with $r_X = 1.14 X^{1/3}$ fm,  $s = 0.9$ fm.

This cross section, Eq.(\ref{mon2}), can be expressed in terms of the recoil
energy of the ordinary atom, $E_R$, and lab velocity, $v$ (i.e.
the velocity in Earth rest frame):
\begin{eqnarray}
{d\sigma \over dE_R} = {\lambda \over E_R^2 v^2}
\label{cs}
\end{eqnarray}
where 
\begin{eqnarray}
\lambda \equiv {2\pi \epsilon^2 \alpha^2 Z^2 Z'^2 \over M_A} \
F_{A}^2 (qr_A) F_{A'}^2 (qr_{A'})
\end{eqnarray}
Note the $1/E_R^2$ dependence. It arises because the dark
matter particles interact electromagnetically (i.e.
via exchange of massless photons). This is
quite unlike the standard WIMP case and therefore
represents a major difference between mirror dark matter
and standard WIMP dark matter.

The interaction rate is
\begin{eqnarray}
{dR \over dE_R} &=& 
\sum_{A'} N_T n_{A'} \int {d\sigma \over dE_R} {f(v,v_E) \over k} |v|
d^3v \nonumber \\
&=& \sum_{A'} N_T n_{A'}
{\lambda \over E_R^2 } \int^{\infty}_{v_{min}
(E_R)} {f(v,v_E) \over k|v|} d^3 v 
\label{55}
\end{eqnarray}
where $N_T$
is the number of target atoms per kg of detector\footnote{
For detectors with more than one target element 
we must work out the
event rate for each element separately and add them up to get the total
event rate.} 
and $f(v,v_E)/k$ is the velocity distribution 
of the mirror element, $A'$, with $v$ being the velocity relative
to the Earth, and $v_E$ is the Earth velocity relative to the
dark matter distribution. The lower velocity limit,
$v_{min} (E_R)$, is obtained from Eq.(\ref{v}), while
the upper limit, $v_{max} = \infty$, because
of $A'$ self interactions (as we will explain in a moment).

The velocity integral in Eq.(\ref{55}),
\begin{eqnarray}
I (E_R) \equiv \int^{\infty}_{v_{min}(E_R)} {f(v,v_E) \over k|v|} d^3 v
\end{eqnarray}
is standard (as it occurs also in the usual
WIMP interpretation) and can easily be evaluated in terms of
error functions assuming
a Maxwellian dark matter distribution\cite{ls},
$f(v,v_E)/k = (\pi v_0^2)^{-3/2} \ exp[-(v+v_E)^2/v_0^2]
$,
\begin{eqnarray}
I(E_R) = {1 \over 2v_0 y}\left[ erf(x+y) - erf(x-y)\right] 
\label{ier}
\end{eqnarray}
where 
\begin{eqnarray}
x \equiv {v_{min} (E_R) \over v_0}, 
\ y \equiv {v_E \over v_0}.
\label{xy}
\end{eqnarray}
For standard non-interacting WIMPs, $v_0$ is expected to be in
the ($90\%$ C.L.) range\cite{koch},
\begin{eqnarray}
170 \ km/s \stackrel{<}{\sim} v_0 \stackrel{<}{\sim} 270\ km/s.
\label{range2}
\end{eqnarray}

In the case of a halo composed of $H', He',$ heavier
mirror elements and dust particles,
there are important differences due to mirror
particle self interactions. 
For example, assuming a number density of $n_{He'} \sim 0.08 \ cm^{-3}$
[c.f. Eq.(\ref{08})]
the mean distance between $He' - He'$ collisions is
$1/(n_{He'} \sigma_{elastic}) \sim 0.03$ light years
(using $\sigma_{elastic} \sim 3\times 10^{-16}\ cm^2$).
One effect of the self interactions is to locally thermally equilibrate
the mirror particles in the halo. 
The $He'$ (and other mirror particles) should be well described by 
a Maxwellian velocity distribution with no cutoff velocity.
[$He'$ do not escape from the halo because of their
self interactions].
A temperature, $T$, common to all the mirror particles 
in the halo can be defined, where
$T = M_{A'}v_0^2/2$ (of course, $T$ will depend on the spatial position).
One effect of this is that $v_0$
should depend on $M_{A'}$ with 
\begin{eqnarray}
v_0 (A') = v_0 (He') \sqrt{M_{He'}/M_{A'}}.
\end{eqnarray}
Thus, knowledge of $v_0$ for $He'$ will fix $v_0$ 
for the other elements.
We will assume that the halo is
dominated by $He'$ (which is suggested by mirror BBN 
arguments\cite{comelli}), with $v_0 \equiv v_0 (He')$ in
the range, Eq.(\ref{range2}).

The Earth motion around the sun produces an annual modulation 
in $y$:
\begin{eqnarray}
y \simeq y_0 + \Delta y \cos \omega (t - t_0)
\end{eqnarray}
where $y_0 = \langle v_E \rangle/v_0$
and $\Delta y = \Delta v_E/v_0$ (for $He'$, $y_0 \approx 1.06$, 
$\Delta y \approx 0.07$). The parameter $t_0$ turns out to be
June 2, and $\omega = 2\pi/T$ (with $T = 1$ year).
Expanding $I(E_R)$ [Eq.\ref{ier}] into a 
Taylor series [making the $y$ dependence
explicit, i.e. $I(E_R,y) \equiv I(E_R)$]:
\begin{eqnarray}
I(E_R,y_0+\Delta y \cos \omega (t-t_0)) = I(E_R,y_0) + \Delta y \cos 
\omega (t-t_0) \left( {\partial I \over \partial
y}\right)_{y=y_0}
\end{eqnarray}
and
\begin{eqnarray}
\left({\partial I \over \partial y}\right)_{y=y_0} = -{I(E_R,y_0) \over y_0} + 
{1 \over \sqrt{\pi} v_0
y_0} \left[ e^{-(x-y_0)^2} + e^{-(x+y_0)^2 }\right]\ .
\end{eqnarray}
The net effect is an interaction rate
\begin{eqnarray}
{dR \over dE_R} = {dR_0 \over dE_R} + 
{dR_1 \over dE_R}
\label{111}
\end{eqnarray}
where
\begin{eqnarray}
{dR_0 \over dE_R} &=& 
\sum_{A'} {N_T n_{A'} \lambda  I(E_R,y_0) 
\over E_R^2}
\nonumber \\
{dR_1 \over dE_R} &=&
\sum_{A'} {N_T n_{A'} \lambda \Delta y \cos \omega (t-t_0) 
\over E_R^2}
\left({\partial I \over \partial y}\right)_{y=y_0}
\ .
\label{above}
\end{eqnarray}
Clearly, galactic mirror atom interactions
will generate an annual modulation, $A \cos \omega (t - t_0)$, 
in the event rate coming from the ${d{R}_1
\over d{E}_R}$ component.

To compare these interaction rates with the experimental measurements,
we must take into account the finite energy resolution 
and quenching factor.
The quenching factor relates the detected energy
($\stackrel{\sim}{E_R}$)
to the actual recoil energy ($E_R$),
\begin{eqnarray}
\stackrel{\sim}{E_R} = q_{A} E_R
\end{eqnarray}
and for the DAMA experiment, $q_{Na}, \ q_I$ have been measured to
be approximately $q_{Na} \simeq 0.30, \ q_I \simeq 0.09$\cite{dama}.
The energy resolution can be accommodated by convolving the rate
with a Gaussian, with $\sigma_{res} \approx 0.16 \stackrel{\sim}{E_R}$
(from figure 3 of Ref.\cite{bern}).

The DAMA collaboration give their results in terms
of the residual rate in the cumulative energy interval
2-6 keV, where they find that
\begin{eqnarray}
A_{exp} = 0.019 \pm 0.003 \ {\rm cpd/kg/keV}
\end{eqnarray}
This number should be compared with the theoretical
expectation:
\begin{eqnarray}
A_{th} = {1 \over 4}\sum_{j=0}^3 A_{th}^j
\end{eqnarray}
where
\begin{eqnarray}
A_{th}^j \equiv \sum_{A=Na,I} \ {1 \over \Delta E} \int^{E_{j}+\Delta E}_{E_j}
\int^{\infty}_{0}
{q_{A}N_T n_{A'} \lambda \Delta y  
\over \stackrel{\sim}{E'}_R^2
\sqrt{2\pi} \sigma_{res}} 
\left({\partial I \over \partial y}\right)_{y=y_0}
e^{{-(\stackrel{\sim}{E}_R - \stackrel{\sim}{E'}_R)^2 \over
2\sigma^2_{res}}} 
\ d\stackrel{\sim}{E'}_R
d\stackrel{\sim}{E}_R
\nonumber \\
.
\label{above4}
\end{eqnarray}
with $E_j = 2.0$ keV $+ \ \Delta E * j \ (j=0, 1, 2,...)$ and 
$\Delta E = 1.0$ keV.

We have numerically studied $A_{th}$. 
We find that the $O'$ contribution to DAMA/NaI
dominates over the $He'$ ($H'$) contribution provided that
$\xi_{O'}/\xi_{He'} \stackrel{>}{\sim} 7 \times 10^{-4}$
($\xi_{O'}/\xi_{H'} \stackrel{>}{\sim} 4 \times
10^{-8}$).
The reason for this is of course clear: the actual threshold recoil 
energy is $6.7$ keV for $A' - Na$ interactions\footnote{
Numerically we find that mirror atom interactions with $Na$ dominate over
$I$ for recoil energies above the 2 keV software threshold.
This reason for this is clear: the threshold velocity is much lower
for interactions with $Na$ which is because a) $Na$ is
a much lighter element than $I$ [c.f. Eq.(\ref{v})]
and b) the quenching factor for $Na$ is 0.3 (c.f. with
0.09 for $I$) which means that the actual recoil threshold
energy is 6.7 keV for $Na$ and 22 keV for $I$.
Also note that the corrections due to the form factor,
which were taken into account using the simple
analytic expression, Eq.(\ref{ff}), are
reasonably small ($\sim 5\%$) for $A' = O'$ but
larger ($\sim 30\%$) for $A'=Fe'$.}, 
which means
that the $v_{min}$ [from Eq.(\ref{v})]
is:
\begin{eqnarray}
v_{min} (H' - Na) & = & 2830 \ km/s 
\nonumber \\
v_{min} (He' - Na) & = & 795 \ km/s 
\nonumber \\
v_{min} (O' - Na) & = & 290 \ km/s 
\nonumber \\
v_{min} (Fe' - Na) & = & 166 \ km/s 
\end{eqnarray}
Because $v_{min} (H'-Na) \gg v_0$, any
$H'$ contribution to the DAMA/NaI signal
is expected to be very tiny and we will neglect it.
The quantity $v_{min} (He'-Na)$ is also quite high which
suppresses the $He'$ contribution relative to $O'$ and $Fe'$.

Interpreting the annual modulation signal in 
terms of $O'$ and $Fe'$, i.e. setting
$A_{th} = A_{exp}$, we find numerically that:
\begin{eqnarray}
|\epsilon | \sqrt{ {\xi_{O'} \over 0.10} +
{\xi_{Fe'} \over 0.026}} 
\simeq 4.8^{+1.0}_{-1.3} \times 10^{-9}
\label{dama55}
\end{eqnarray}
where the errors denote a 3 sigma allowed range (corresponding
to $0.010 < A_{exp} < 0.028$). The best fit region will
also be affected by systematic uncertainties in 
the quenching factors, form factors and astrophysical
uncertainties [e.g. uncertainties in $v_0 (A')$].
These uncertainties will increase the possible
parameter range, however a detailed investigation
of these effects we leave for the future.

Because of the different masses of the two
components, $O'$ and $Fe'$, their relative
contributions can potentially be determined by
the differential recoil energy spectrum, $A^j$.
In {\bf figure 1} we examine
the representative possibilities a) DAMA signal is dominated
by $O'$ [i.e. $\xi_{O'} = 0.10$, $\xi_{A'} = 0$ for $A' \neq O'$],
b) DAMA signal is dominated by $Fe'$ [i.e.
$\xi_{Fe'} = 0.026$, $\xi_{A'} = 0$ for $A' \neq Fe'$].
The case where $A_{th}$ is made up of approximately equal
contributions from both $O'$ and $Fe'$,
corresponding to $\xi_{O'} \simeq 4\xi_{Fe'} = 0.05$ is also given.
However, again we point out that a careful study
of systematic uncertainties will be necessary before
any definite conclusions can be made about the ratio
of $\xi_{O'}/\xi_{Fe'}$.

Besides the annual modulation effect, DAMA/NaI has also
measured the absolute event rate. This rate will contain the signal
[$dR_0/dE_R$, Eq.(\ref{above}), convolved with a Gaussian to
incorporate the detector resolution]
plus any background contribution. An interesting point
is that the cross section, Eq.(\ref{cs}), rises
sharply ($\propto 1/E_R^2$) at low $E_R$, and this effect
may show up in the data. [In any case, we should check that
our absolute rate from the signal does not exceed the measured
absolute rate].
In {\bf figure 2} we plot the absolute rate 
with parameters fixed by the 
annual modulation signal, Eq.(\ref{dama55}).
Also plotted is the measured rate obtained from
Ref.\cite{dama5}. Interestingly, the data does indeed 
show a sharp rise at low $E_R$ which is compatible
with the parameters suggested by the annual modulation
effect. The {\it shape} of the measured rate at low $E_R$ 
is nicely fitted by
both $O'$ and $Fe'$ dark matter, but the
normalization prefers $O'$ over $Fe'$ dark matter. However
possible small systematic uncertainties such as calibration
errors may be present: a 0.1-0.2 keVee calibration error
would be enough to allow $Fe'$ to fit the data at low
$E_R$. 

Implicit in our analysis is that the mirror atoms can
reach the DAMA detector from all directions, without getting
stopped in the Earth. 
The stopping distance of a mirror
atom, $A'$ (of energy $E' = {1 \over 2}M_{A'} v^2$)
in ordinary matter (of atomic number density $n = \rho/M_A$) 
can easily be evaluated from:
\begin{eqnarray}
{dE' \over dx} &=& -{\rho \over M_A}\int E_R {d\sigma \over dE_R} dE_R
\nonumber \\
&=& {-\rho \pi M_{A'} \epsilon^2 \alpha^2 Z^2 Z'^2 \ln \left({E_R^{max} \over
E_R^{min}}\right) \over M_A^2 E' }
\label{rrt}
\end{eqnarray}
where $E_R^{max}$ can be obtained from Eq.(\ref{v}) and
$E_R^{min} = 1/(2r_0^2 M_A)$ (due to atomic screening).
[Explicitly, $\ln\left({E_R^{max}\over E_R^{min}}\right) \approx 10$].
Eq.(\ref{rrt}) can be solved to give the energy of the mirror
atom after travelling a distance $x$ through ordinary matter:
\begin{eqnarray}
E'(x) = E'(0) \sqrt{1 - {x \over L}}
\end{eqnarray}
where $L$ is the stopping distance:
\begin{eqnarray}
L &\simeq & {M_A^2 M_{A'} v_i^4 \over 8\pi \rho \epsilon^2
\alpha^2 Z^2 Z'^2 10
}
\nonumber \\
&\approx &
10^5 \left( {10^{-8} \over \epsilon}\right)^2 \left( {v_i \over 400 \
km/s}\right)^4 \left({5 \ g/cm^3 \over \rho}\right)\left( {2 \over
Z'}\right)
\ {\rm km}
\label{stop}
\end{eqnarray}
where $v_i$ is the initial velocity of the mirror atom.
The stopping distance in earth for $He'$, $O'$
and $Fe'$
can easily be obtained from the above
equation, giving:
\begin{eqnarray}
& L(He') & \stackrel{>}{\sim} 10^7\ km \ {\rm for} \ |\epsilon | = 4 \times
10^{-9}, \ v_i \ge v_{min} (He') \simeq 795\ km/s, 
\nonumber \\
& L(O') & \stackrel{>}{\sim} 5\times 10^4\ km \ {\rm for} \ |\epsilon
| = 4 \times
10^{-9}, \ v_i \ge v_{min} (O') \simeq 290\ km/s. 
\nonumber \\
& L(Fe') & \stackrel{>}{\sim} 3 \times 10^3\ km \ {\rm for} \ |\epsilon | = 
4 \times
10^{-9}, \ v_i \stackrel{>}{\sim} 200 \ km/s. 
\end{eqnarray}
Since $L(He'), L(O')$ are much larger than the Earth's diameter,
the retarding effect of the Earth is relatively small and
no large diurnal effect is expected (in agreement
with DAMA observations\cite{diurnal}).
Mirror iron, may lead to a possibly large diurnal effect.
However, dark matter detection experiments depend on $|\epsilon |
\sqrt{\xi_{A'}}$ while the stopping distance in earth, depends
just on $|\epsilon |$. 
The significant uncertainty in the size of $\xi_{A'}$ implies corresponding
uncertainty in $\epsilon$ and hence $L(A')$.
It is therefore still possible
for the DAMA/NaI signal to be dominated by the $Fe'$ component,
without leading to any significant diurnal effect.
Note that experiments with a lower threshold (and hence
lower value of $v_{min}$) will have a much greater sensitivity
to the diurnal effect, so this effect may show up
in future experiments.

Let us now consider implications of this interpretation
of the DAMA signal for other experiments.
The CDMS/Ge experiment\cite{cdms} has searched for nuclear recoils
due to WIMP-Ge elastic scattering. This experiment
has a threshold energy of $10$ keV and the quenching
factor is {\it assumed} (but not measured!) to be 1. 
This experiment finds just 4 events
satisfying their cuts with $10 \ keV < E < 20 \  keV$
for their exposure of 10.6 kg-day. However
because the target consists of the relatively heavy element,
$Ge$, and the threshold is relatively high, 10 keV, the
sensitivity of the CDMS experiment to light mirror elements,
is completely negligible. Assuming $|\epsilon |
\sqrt{\xi_{O'}/0.10} = 4.8\times 10^{-9}$ (as suggested from the
DAMA/NaI experiment, if $O'$ dominates the rate),
we find numerically that the number of
$O'$ induced events (above the 10 keV CDMS threshold) is much less than
1 for their exposure of 10.6 kg-day.

If there happens to be a significant $Fe'$ component, then this
may potentially be constrained by CDMS/Ge experiment.
In the case where $Fe'$ dominates the DAMA/NaI
experiment, then $|\epsilon |\sqrt{\xi_{Fe'}/0.026} = 4.8\times 10^{-9}$.
Numerically, we find that this
implies 26 events in the $10\ keV < E < 20\ keV$
range for CDMS, for their 10.6 kg-day exposure (c.f. just 4 detected
events). The low rate obtained by CDMS experiment
suggests that $Fe'$ does not dominate over $O'$.
However given possible experimental uncertainties,
the case of $Fe'$ dominance is probably not completely excluded.
For example, the quenching factor may turn out to be somewhat less
than 1. For example, a value of 0.6 would reduce the expected number of
events from 26 down to 5 events which is consistent with the
data.

Clearly experiments with a lower threshold than DAMA/NaI might
potentially provide more stringent constraints.
The only experiment with a lower threshold than DAMA/NaI
is the CRESST/Sapphire experiment\cite{cresst}.
That experiment uses 262 g sapphire crystals ($Al_2 O_3$) as the target
medium with a low detection threshold of
$E_R (threshold) = 0.6$ keV. 
These features make CRESST/Sapphire 
particularly sensitive to low mass dark
matter particles such as $He', O'$ (and even $Fe'$). 
Unfortunately, the CRESST experiment does not 
have enough statistics to be sensitive to the annual
modulation due to the Earth's motion around the sun, nevertheless
the shape and normalization of the measured energy spectrum 
provide useful information\footnote{
Because of the low threshold, the CRESST experiment
might be sensitive to the diurnal effect and this could
even show up in the existing data.}.
We now study in detail the implications of mirror matter-type
dark matter for this experiment.

In this
experiment the quenching factor is assumed to be approximately
equal to 1 (however, again this has not been specifically 
measured)\cite{cresst}.
As with the DAMA/NaI experiment,
the recoil spectrum, Eq.(\ref{55}), needs be convolved
with a Gaussian curve (with $\sigma_{res} \simeq 0.4247\Delta
E_{res}$\footnote{
Note that $\sigma_{res} = {\Delta E_{res} \over \sqrt{8\ln 2}} 
\simeq 0.4247\Delta E_{res}$
implies a FWHM of $\Delta E_{res}$ for the Gaussian curve,
which is the CRESST prescription\cite{cresst}.
In Ref.\cite{cresst}, two values of $\Delta E_{res}$ are
discussed,
$\Delta E_{res} \approx 0.2$ keV (from an internal
calibration source) and $\Delta E_{res} \approx
0.5$ keV (from possible contamination with $^{55}Fe$).
In our numerical work we have used the former
value (unless otherwise stated). 
Using $\Delta E_{res} = 0.5$ keV would lead to 
$|\epsilon | \sqrt{\xi_{A'}}$ values smaller by about
$20 \%$.} 
)
in order to take into account the finite 
energy resolution of the detector, 
\begin{eqnarray}
{d\stackrel{\sim}{R} \over dE_R} =
\int^{\infty}_{0} {dR(E'_R) \over dE'_R}
\ {1 \over \sqrt{2\pi} \sigma_{res}}
\ e^{{-(E_R - E'_R)^2 \over
2\sigma^2_{res}}} 
\ dE'_R
\label{above2}
\end{eqnarray}
The CRESST collaboration present their results in terms of the 
quantity,
\begin{eqnarray}
C_j \equiv {1 \over \Delta E} \int^{E_j + \Delta E}_{E_j}
{d\stackrel{\sim}{R} \over dE_R} 
\ dE_R
\label{x3}
\end{eqnarray}
where $E_j = 0.6 \ {\rm keV} + \Delta E * j$ ($j=0,1,2,...$) and
$\Delta E = 0.2$ keV. 
We have numerically studied $C_j$ for various cases.
In {\bf figure 3} we plot the expected value
for $C_j$ for the best fit values of $|\epsilon |\sqrt{\xi_{A'}}$
assuming the DAMA/NaI rate is dominated by a) $O'$, b) $Fe'$
and c) 50-50 $O'$, $Fe'$ mixture.
[Recall, these are the same three cases which were fitted
to the DAMA/NaI annual modulation signal and were plotted in 
figure 1].
While the shape of the CRESST/Sapphire data
(obtained from figure 10 of Ref.\cite{cresst})
is reasonably consistent with the expected shape from $A'$ 
interactions, the normalizable is roughly a factor of 2 too
high. This may be due to systematic uncertainties which
we illustrate in {\bf figure 4}.
In this figure, the CRESST quenching factor is taken to be $0.7$
instead of the assumed value of 1.0.
(Similar results occur if there happens to be 
a small energy calibration uncertainty of $\sim 0.2$
keV).
Figure 4 clearly demonstrates the rather nice fit
of $O', Fe'$ dark matter
to the shape and normalization 
of the CRESST data
(after allowing for reasonable systematic uncertainties).

Given the rather nice fit of the shape and normalization of the
CRESST data (within
reasonable systematic uncertainties)
to the expectations of $Fe', O'$ dark matter from the
DAMA/NaI fit, it is clearly very tempting to suppose
that the CRESST data may be mostly signal with very little
background component.
On the other hand,
the CRESST collaboration\cite{cresst} have argued that their
data is most likely background because of the rate
of coincidence events. This argument required
the background to be due to single particle
interactions and isotropic which it may not be.


Finally, note that the CRESST/Sapphire experiment 
is much more sensitive to $H', He'$ than
the DAMA/NaI experiment.
Assuming a pure $He'$ halo, i.e. $\xi_{A'} = 0$
for $A' \neq He'$, 
we find that the CRESST data suggest:
\begin{eqnarray}
|\epsilon | \sqrt{ {\xi_{He'} \over 0.5} } \approx 4 \times
10^{-9}
\end{eqnarray}
In this pure $He'$ halo limit,
the (CRESST) value for 
$|\epsilon |\sqrt{\xi_{He'}}$, above, is not consistent with the
value from DAMA/NaI [Eq.(\ref{dama55})].
Thus $He'$ cannot dominate the rate for DAMA or CRESST.
This suggests that:
$\xi_{O'} \stackrel{>}{\sim} 0.2 \xi_{He'}$ and/or 
$\xi_{Fe'}  \stackrel{>}{\sim} 0.04 \xi_{He'}$. 
Clearly this constraint is significant, but
nevertheless, still allows $He'$ to be the dominate halo 
dark matter component\footnote{
Note that the CRESST/Sapphire experiment does not
put significant limits on the $H'$ component of
the dark matter halo.
Numerically, we find that $He'$ dominates over $H'$
provided that $\xi_{H'} \stackrel{<}{\sim} 15\xi_{He'}$ 
which is not a very stringent condition.}.

Assuming that DAMA and CRESST have detected
galactic mirror matter-type dark matter, then this suggests an
$|\epsilon |$ value of around $10^{-8}-10^{-9}$. 
Previous work (see Ref.\cite{fm} and references there-in) 
looking at various solar system implications
of mirror matter has identified a similar 
but somewhat larger range for $\epsilon$, Eq.(\ref{range}).
This information is summarized in {\bf figure 5}.
Also shown is the experimental
bound\cite{ortho,fg}, $|\epsilon | \stackrel{<}{\sim} 5\times 10^{-7}$ coming
from recent orthopositronium lifetime measurements\cite{gid} 
and also the
limit suggested from BBN\cite{cg}.

Let us also mention that if $|\epsilon | \sim 5\times 10^{-9}$,
there will be interesting terrestrial effects of mirror matter.
Fragments (of size $R$) of impacting mirror matter space bodies can remain
on/near the Earth's surface provided that\cite{cent}
\begin{eqnarray}
R \stackrel{<}{\sim} 5 \left( {|\epsilon | \over  5 \times
10^{-9}}\right) \ {\rm cm}.
\end{eqnarray}
Such fragments can potentially be detected and extracted with a
centrifuge\cite{cent}. If mirror matter fragments become completely embedded
within
ordinary matter (which is necessarily the case for $\epsilon < 0$) 
then the fragments will thermally equilibrate with the
ordinary matter environment.
The observational effect of this is to cool the surrounding
ordinary matter, as heat is transferred to the mirror
body and radiated away into mirror photons\cite{thermal}.
Finally, even tiny solar system mirror dust particles can
lead to observable effects.
These particles impact with the Earth with velocity
in the range $11 \ {\rm km/s}\stackrel{<}{\sim} v \stackrel{<}{\sim} 70
\ {\rm km/s}$ and can be detected
in suitably designed surface experiments\cite{last}
such as the
St. Petersburg experiment\cite{drob}.

In conclusion, we have pointed out that the DAMA/NaI, CRESST/Sapphire
and other dark matter experiments
are sensitive to mirror matter-type dark matter.
Furthermore, the annual modulation signal
obtained by the DAMA/NaI experiment can be explained by mirror
matter-type dark matter
for $|\epsilon | \sim 5 \times 10^{-9}$. This explanation of
the DAMA signal is 
supported by DAMA's absolute rate measurement as well
as by the size and shape of the CRESST data. Furthermore
this explanation is not
in conflict with CDMS or any of the other dark matter
experiments because of their higher thresholds. 
The  
$\epsilon $ value suggested by the DAMA/NaI experiment
is consistent with the value obtained from
various solar system anomalies including the Pioneer spacecraft
anomaly, anomalous meteorite events and lack of small
craters on the asteroid Eros.
It is also consistent with standard BBN.

\vskip 1cm
\noindent
{\bf \large Acknowledgements:}
The author would like to thank Profs. R. Bernabei, R. Cerulli, F. Probst
for patiently helping me understand some
details regarding their experiments. 
The author also thanks R. Bernabei and R. Volkas for very valuable
comments on a draft of this paper.

\newpage
\centerline{\epsfig{file=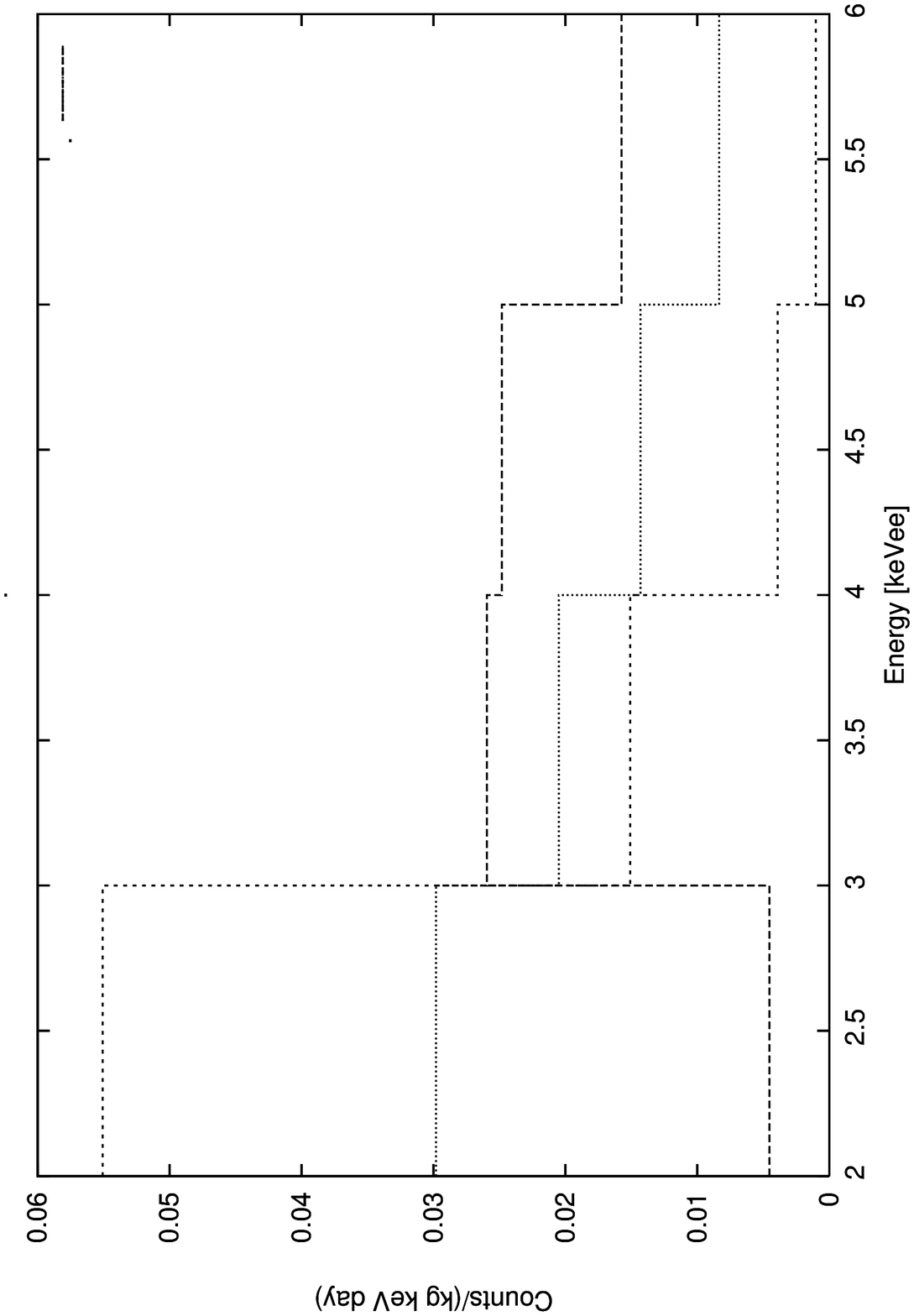, width=10.0cm}}
\vskip 0.5cm
\noindent
{\small Figure 1: 
$A_{th}^j$ (as defined in text) corresponding to the DAMA experiment 
for $O',\ Fe'$ dark matter
(for $v_0 = 230$ km/s),
with parameters
a) $|\epsilon | \sqrt{\xi_{O'}/0.10} = 4.8 \times 10^{-9}$, 
$\xi_{A'} = 0$ for $A' \neq O'$
(short-dashed line), 
b) $|\epsilon | \sqrt{\xi_{Fe'}/0.026} = 4.8 \times 10^{-9}$,
$\xi_{A'} = 0$ for $A' \neq Fe'$
(long-dashed line), 
c) $|\epsilon | = 4.8 \times 10^{-9}$
with $\xi_{O'} = 4\xi_{Fe'}= 0.05$
(dotted line).
In all three cases the differential rate in the 2-6 keV 
window agrees with the experimental value: 
${1 \over 4}\sum_{j=0}^{3} A^j_{th} \simeq A_{exp}$ and the effect
for $\stackrel{\sim}{E}_R > 6$ keV is negligible.
}

\newpage
\centerline{\epsfig{file=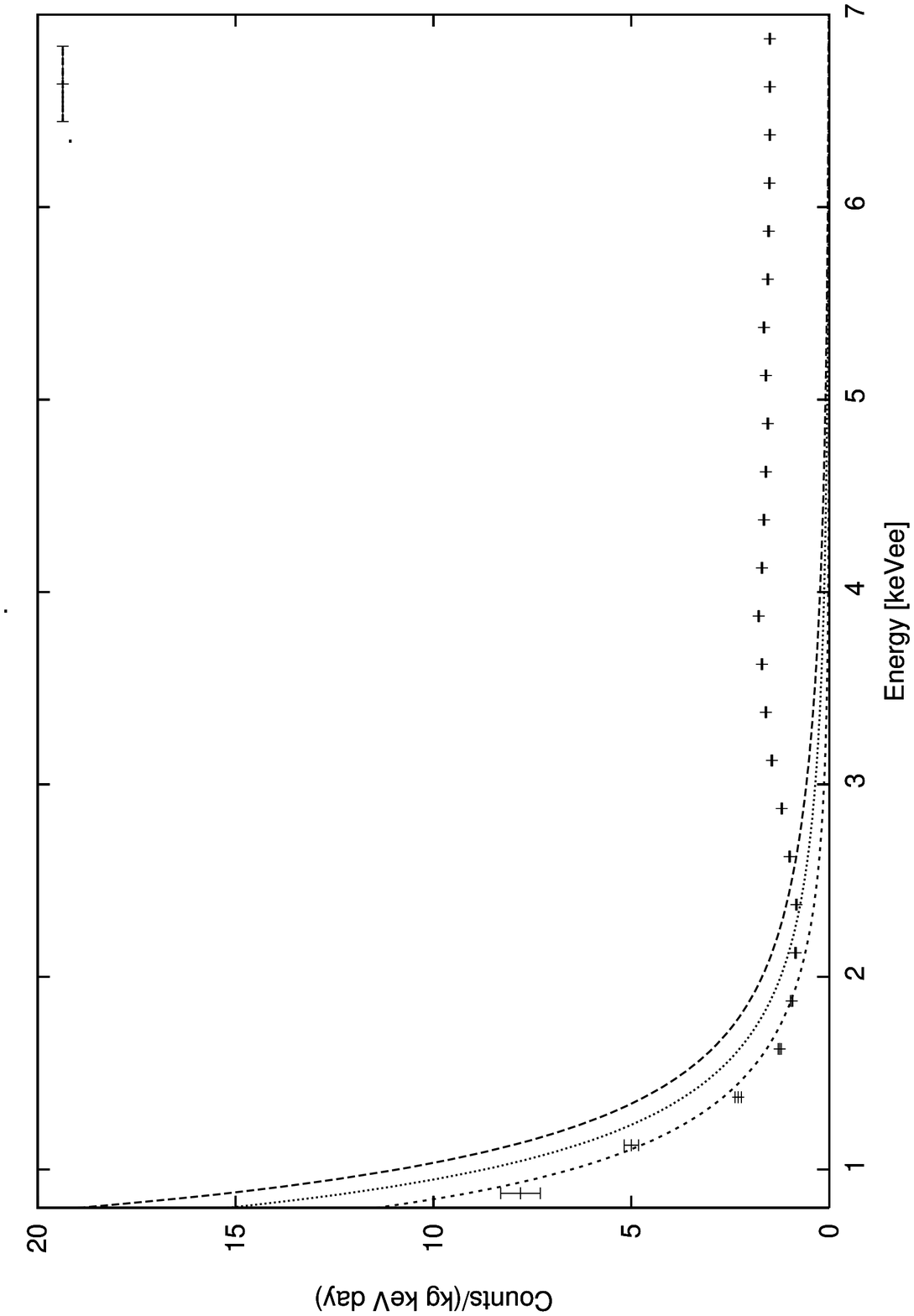, width=10.0cm}}
\vskip 0.5cm
\noindent
{\small Figure 2:
The absolute event rate for the DAMA/NaI experiment
for $O',\ Fe'$ dark matter
(for $v_0 = 230$ km/s).
The parameters are given by the fit to the
DAMA/NaI annual modulation effect, where
we take the same three representative cases
as figure 1:
a) $|\epsilon | \sqrt{\xi_{O'}/0.10} = 4.8 \times 10^{-9}$,
$\xi_{A'} = 0$ for $A' \neq O'$ (short-dashed line)
b) $|\epsilon | \sqrt{\xi_{Fe'}/0.026} = 4.8 \times 10^{-9}$,
$\xi_{A'} = 0$ for $A' \neq Fe'$ 
(long-dashed line) and c)
$|\epsilon | = 4.8 \times 10^{-9}$
with $\xi_{O'} = 4\xi_{Fe'}= 0.05$
(dotted line).
Also shown is the DAMA/NaI data obtained from
Ref.\cite{dama5}.
}

\newpage
\vskip 0.5cm
\centerline{\epsfig{file=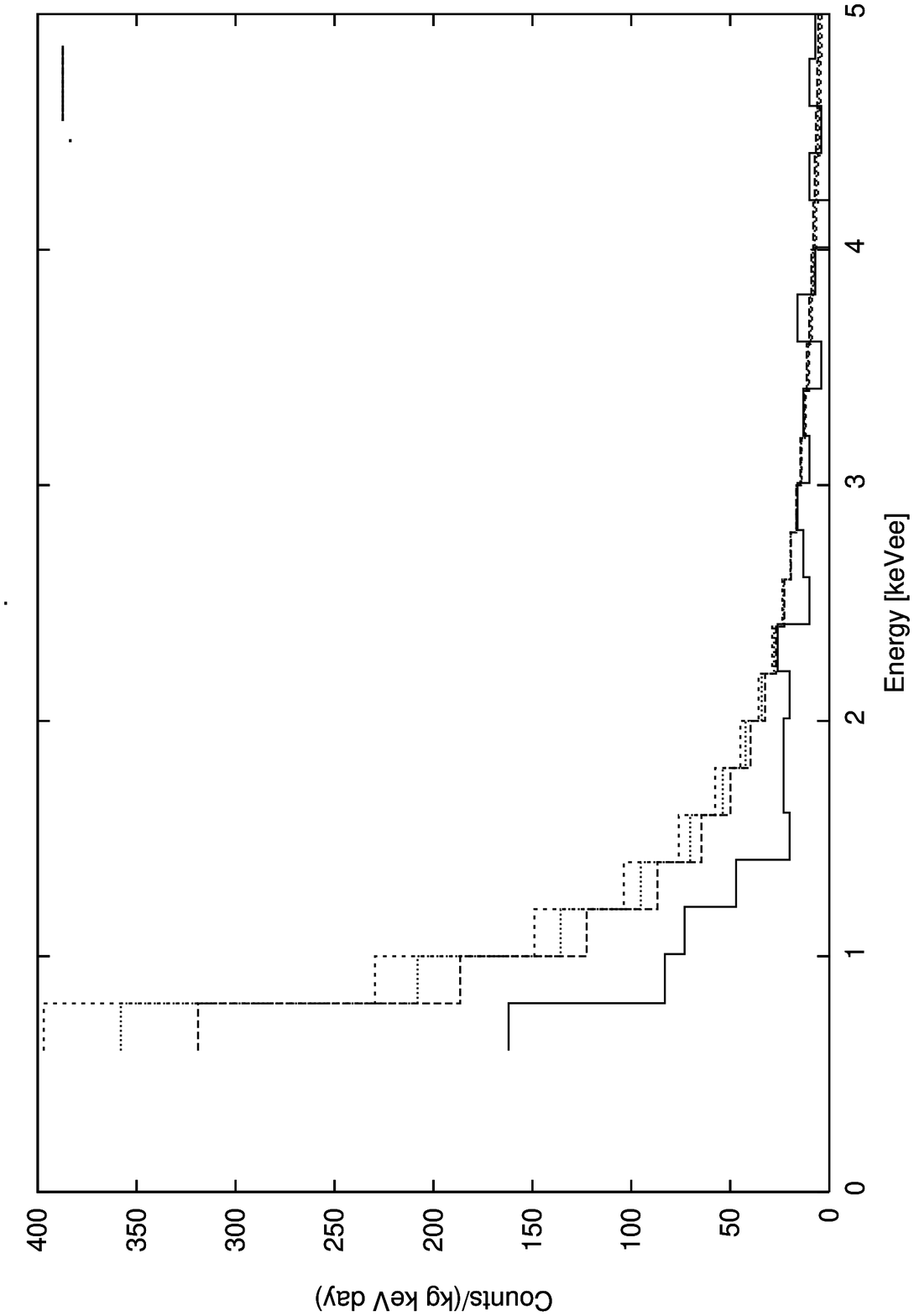, width=10.0cm}}
\vskip 0.5cm
\noindent
{\small Figure 3: Expectation for the CRESST experiment
for $O', \ Fe'$ dark matter with $v_0 = 230$ km/s and 
a) $|\epsilon | \sqrt{\xi_{O'}/0.10} = 4.8\times 10^{-9}$
(short-dashed line), 
b) $|\epsilon | \sqrt{\xi_{Fe'}/0.026} = 4.8\times 10^{-9}$ 
(long-dashed line), 
c) $|\epsilon | = 4.8\times 10^{-9}$ 
with $\xi_{O'} = 4\xi_{Fe'} = 0.05$
(dotted line).
Also shown (solid line) is the CRESST data obtained 
from figure 10 of Ref.\cite{cresst}.
Note that the statistical errors in the data 
are $\sim 15-30\%$ for $E_R/keV = 0.6 - 2.0$ and 
$> 30\%$ for $E_R > 2.0$ keV.  }

\newpage
\vskip 0.5cm
\centerline{\epsfig{file=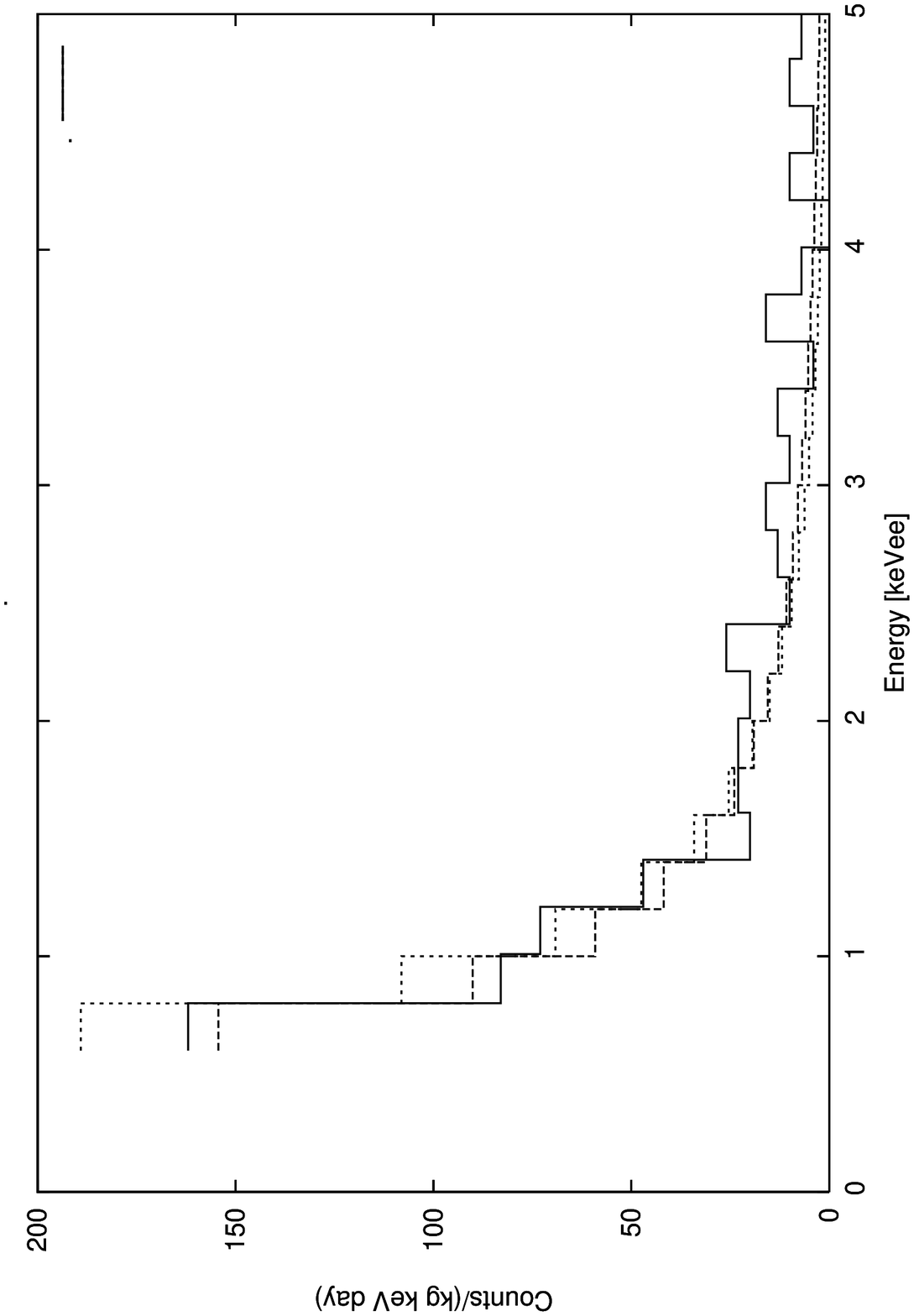, width=10.0cm}}
\vskip 0.5cm
\noindent
{\small Figure 4: Expectation for the CRESST experiment for $Fe'$
dark matter with 
$v_0 = 230$ km/s and
a) $|\epsilon | \sqrt{\xi_{O'}/0.10} = 4.0\times
10^{-9}$ (short-dashed-line), 
b) $|\epsilon | \sqrt{\xi_{Fe'}/0.026} = 4.0\times
10^{-9}$ (long-dashed-line). 
In both cases, a 
CRESST quenching
factor of $q = 0.7$ (instead of 1) has been assumed. 
Also shown (solid line) is the CRESST data. 
}

\newpage
\vskip 0.5cm
\centerline{\epsfig{file=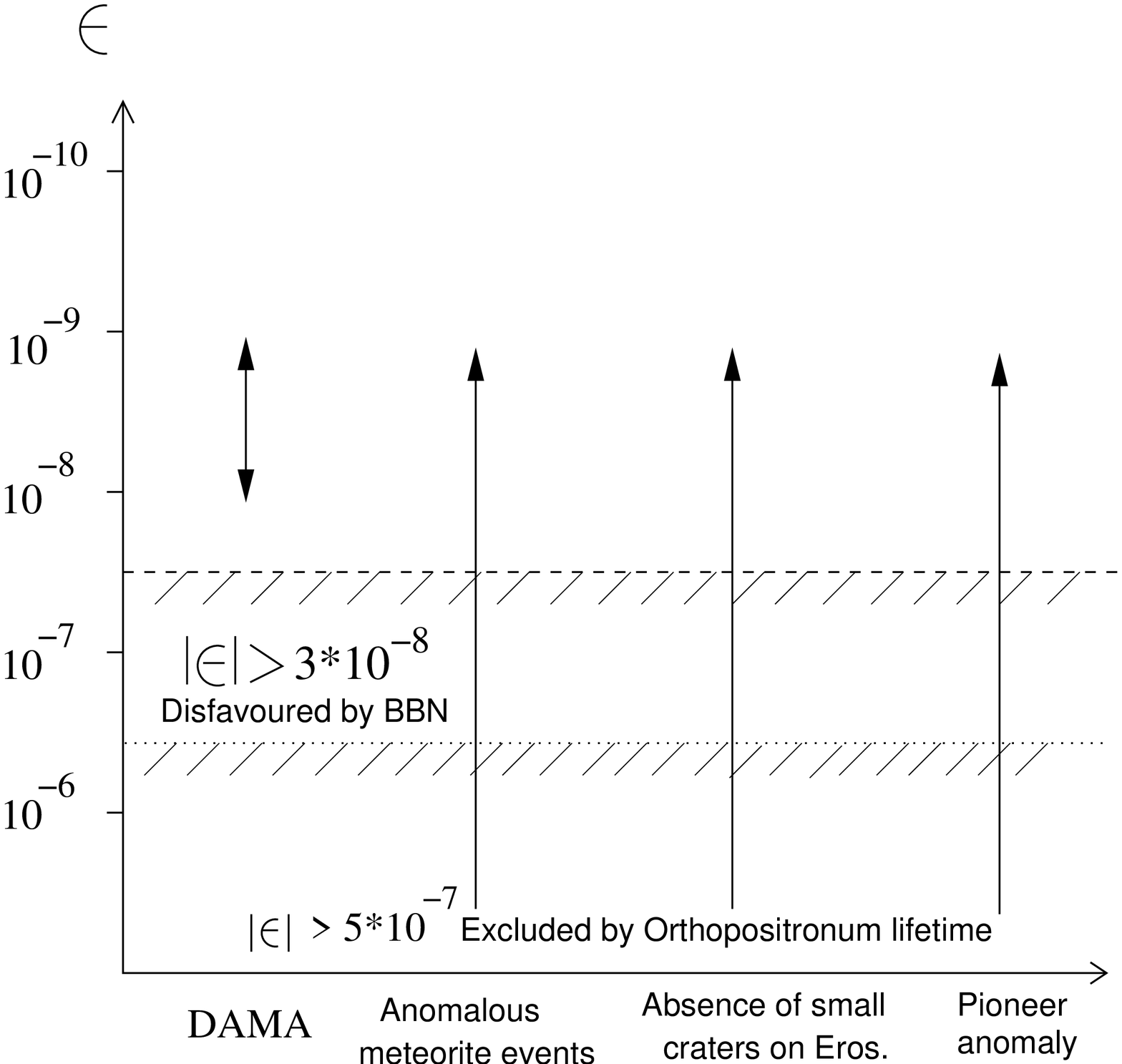, width=10.0cm}}
\vskip 0.5cm
\noindent
{\small Figure 5: Favoured range of $\epsilon$ from various
experiments/observations.} 


\end{document}